\documentclass[prd,aps,secnumarabic,showpacs]{revtex4}
\usepackage{amsmath}
\usepackage{slashed}
\usepackage{dsfont}
\usepackage{graphicx}
\usepackage{amssymb}

\newcommand{\brac}[1]{\left( #1 \right)}
\renewcommand{\thefootnote}{\fnsymbol{footnote}}

\begin{document}
\title{Muon decay in a linearly polarized laser field}
\renewcommand{\thefootnote}{\fnsymbol{footnote}}
\author{Arsham Farzinnia$^{1,}$\footnote{Electronic address: farzinni@msu.edu}, 
Duane A. Dicus$^{2,}$\footnote{Electronic address: dicus@physics.utexas.edu},
Wayne W. Repko$^{1,}$\footnote{Electronic address: repko@pa.msu.edu},
and Todd M. Tinsley$^{3,}$\footnote{Electronic address: tinsley@hendrix.edu}}
\affiliation{$^1$Department of Physics and Astronomy, Michigan State University, East Lansing MI 48824  \\
$^2$Physics Department, University of Texas, Austin, TX 78712  \\
$^3$Department of Physics, Hendrix College, Conway AR 72034}
\renewcommand{\thefootnote}{\arabic{footnote}}

\date{\today}

\begin{abstract}
In a previous paper, we showed that the decay rate of a muon is only slightly affected by the presence of a circularly polarized laser and we gave an analytic expression for the correction. In this paper, we present the analytical result for the case of a linearly polarized laser. Again the effect of the laser is small.
\end{abstract}
\maketitle

\section{Introduction}

Previously, some attempts have been made to find the change in the decay rate of the muon whenever a strong laser field is present. Liu, Li, and Berakdar \cite{LLB} (LLB) tried to calculate this change for a strong linearly polarized laser, using an approximated electron wavefunction combined with numerical calculations. They found a large modification of the lifetime, as much as an order of magnitude. Narozhny and Fedotov challenged this result, arguing a small modification of the lifetime through a brief calculation \cite{rus,LLB2}.

Recently, we showed the full analytical calculations of the decay rate of the muon in the presence of a strong circularly polarized laser field, finding only small (explicit) corrections to the unperturbed decay rate \cite{us}.

Although one might not intuitively expect major alterations of this conclusion once a laser with a different polarization is used, it proves to be worth going through the calculations for a strong linearly polarized laser. As we will show, this calculation is much more complicated and tedious. The electron wavefunction will involve two different exponents of the sine function, resulting in triple summations of the Bessel functions; in contrast with the circularly polarized case, where we had only one exponent of the sine function, and consequently, only one summation of the Bessel functions \cite{us}. Also, this calculation will settle any possible disagreement among the community regarding the difference in laser polarization, as well as providing a reference for the future work.

\section{Theory}

Muon decay is represented by
\begin{equation}\label{decay}
\mu^-(P)\,\longrightarrow\,e^-(p)+\bar{\nu_e}(q_1)+\nu_{\mu}(q_2)\;,
\end{equation}
where the arguments label the associated momenta. The wavefunction of the relativistic electron in a electromagnetic field can be found by solving the Dirac equation with the electromagnetic potential term present. It was first presented by D. M. Volkov in 1935 \cite{volk, V2, BLP}
\begin{equation}\label{dirac}
\left( i\slashed{\partial} - e\slashed{A} - m \right)\psi_e (x) = 0 \; .
\end{equation}
For a linear polarization, $A^{\mu}$ is given by
\begin{eqnarray}\label{A}
A^{\mu}(x)\, &=& \,a^{\mu}\cos\,k\cdot\,x \\
a^{\mu} \, &=& \, \brac{0, \boldsymbol{\mathcal{E}} / \omega}
\end{eqnarray}
with $\boldsymbol{\mathcal{E}}$ the amplitude of the electric field; $\textbf{k} \cdot \boldsymbol{\mathcal{E}} = 0$. We choose the photons to propagate along the z-axis, $k^{\mu}\,=\,(\omega,0,0,\omega)$, and the Volkov solution gives
\begin{equation}\label{wavef}
\psi_e (x) = \left( 1 + \frac{e\slashed{k} \slashed{A}}{2p \cdot k} \right) u(p) \, e^{-iq \cdot x \, + \, i \frac{e^2 a^2}{8p \cdot k} \sin \, 2k \cdot \, x } \, e^{-i \frac{e \, p \cdot a}{p \cdot k} \sin\, k \cdot \, x}
\end{equation}
with the electron's effective momentum and mass
\begin{equation}\label{eff}
q^{\mu} = p^{\mu} - \frac{e^2a^2}{4p \cdot \, k}k^{\mu} \; ,  \qquad \qquad m^2 = m_0^2 - \frac{e^2a^2}{2}
\end{equation}
where $m_0$ is the rest mass of the unaffected electron, $m_0=0.511$ MeV. Also, note that $q\cdot\,k\,=\,p\cdot\,k$.

Following the standard $S-$matrix theory, the matrix element for the reaction, $\mathcal{M}_{fi}$, may be extracted from
\begin{equation}\label{s_fi}
S_{fi} = -i \frac{G}{\sqrt{2}} \int \big[ \bar{\psi}_{\nu_{\mu}} \gamma_{\lambda} \brac{1-\gamma_5} \psi_{\mu} \big] \big[ \bar{\psi}_{e} \gamma^{\lambda} \brac{1-\gamma_5} \psi_{\nu_e} \big] d^4 x \equiv -i (2\pi)^4 \delta^{(4)} \brac{p_f - p_i} \mathcal{M}_{fi} \; .
\end{equation}
In order to evaluate the left hand side of \eqref{s_fi}, the following generating function for Bessel functions from {\bf 9.1.41} of \cite{AS}
\begin{equation}\label{gf}
e^{\frac{1}{2}z(t-1/t)}\,=\,\sum_{n=-\infty}^{\infty}t^{n}J_{n}(z)
\end{equation}
may be used to deduce the following (general) relations
\begin{equation}\label{rel}
\begin{Bmatrix} 
1\\ 
\cos z \\ 
\sin z
\end{Bmatrix}
\times e^{-i \, \xi \sin \brac{z - \phi_0} }
= \underset{n=-\infty}{\overset{\infty}{\sum}} e^{-inz} \times
\begin{Bmatrix} 
J_n (\xi) e^{in\phi_0}\\ 
\frac{1}{2} \brac{J_{n+1} (\xi) e^{i(n+1)\phi_0} + J_{n-1} (\xi) e^{i(n-1)\phi_0}} \\ 
\frac{1}{2i} \brac{J_{n+1} (\xi) e^{i(n+1)\phi_0} - J_{n-1} (\xi) e^{i(n-1)\phi_0}}
\end{Bmatrix} \; .
\end{equation}
Using \eqref{rel}, and after some algebra, from \eqref{s_fi} we find the matrix element for a given value of $\ell$ and $L$ (one index per exponent in \eqref{wavef})
\begin{equation}\label{m_Ll}
\mathcal{M}_{L,\ell} = \frac{G}{\sqrt{2}} \, \bar{u}(q_2) \gamma_{\lambda} \brac{1-\gamma_5} u(P) \bar{u}(p)\big[ \Delta_0 + \Delta_1 \slashed{a} \slashed{k} \big] \gamma^{\lambda} \brac{1-\gamma_5} v(q_1)
\end{equation}
where
\begin{align}
& \Delta_0 = J_L (B) J_{\ell} (D) \label{del0} \\
& \Delta_1 = \frac{e}{4p \cdot k} J_L (B) \big[ J_{\ell +1} (D) + J_{\ell -1} (D)\big] \label{del1}
\end{align}
and
\begin{equation}\label{DB}
D = -\frac{e \, p \cdot \, a}{p \cdot \, k} \; , \qquad \qquad  B = \frac{e^2 a^2}{8p \cdot \, k} \; .
\end{equation}
The momentum conservation is then
\begin{equation} \label{mom.con}
P^{\mu}+\brac{2L+\ell}k^{\mu}\,=\,q^{\mu}+q_1^{\mu}+q_2^{\mu} \; .
\end{equation}
Note that $L$ has a prefactor 2, coming from the argument of sine in \eqref{wavef}. Following the standard procedure of determining the decay rate, described for example in \cite{Todd}, we see that once \eqref{s_fi} is squared, there are four summation indices, $L$, $L'$, $\ell$, and $\ell'$. One of the two delta functions, however, takes care of one of the summation indices
\begin{equation}\label{l'}
\ell' = 2\brac{L-L'} + \ell \; ,
\end{equation}
hence, we are left with three summation indices. The total decay rate is then given \cite{Todd} by
\begin{equation}\label{gamma}
\Gamma = \sum_{L, L', \ell} \Gamma_{L, L', \ell}
\end{equation}
where for each set of indices
\begin{equation}\label{partgamma}
\Gamma_{L,L',\ell} = \brac{ \prod_{f} \int \frac{d^3 p_f}{2E_f(2 \pi)^3}} \frac{\brac{2 \pi}^4}{2E_{\mu}} \delta^{(4)} \brac{P-q-q_1-q_2+\brac{2L+\ell} k} \, \frac{1}{2} \underset{s_{\mu},s_{e},s_{\nu_{\mu}},s_{\nu_{e}}}{\sum}  \mathcal{M}_{L',\,2(L-L')+\ell}^* \, \mathcal{M}_{L,\ell} \; .
\end{equation}
Using FORM \cite{form}, all the traces in \eqref{partgamma} can be evaluated, and we may perform the integration over the neutrino momenta by using the well-known relation
\begin{equation}\label{neuint}
\int\frac{d^3q_1}{2q_1^0}\frac{d^3q_2}{2q_2^0}\delta^{4}(Q-q_1-q_2)q_1^{\alpha}q_2^{\beta}\,=\, \frac{\pi}{24}\big(Q^2\,g^{\alpha\beta}+2Q^{\alpha}Q^{\beta}\big)\Theta(Q^2)\; ,
\end{equation}
with $Q = P - q + \brac{2L + \ell}k$. These then reduce \eqref{partgamma} -- in muon rest-frame -- to
\begin{equation}\label{GLL'l}
\Gamma_{L,L',\ell}\,=\,\frac{1}{3072\pi^3M}\int\,dE\,|{\bf q}|\int\,dz\,\Theta(Q^2)\,T_{L,L',\ell}
\end{equation}
where $M$ is the muon rest mass, and $E$ is $q^0$. The object $T_{L,L',\ell}$ is the square of the matrix element \eqref{m_Ll}, summed over the spin and integrated over the neutrino momenta, with relation \eqref{l'} implied (see APPENDIX for the explicit expression.)

Turning off the laser corresponds to $a=0$, which in turn, from \eqref{DB}, sets $D$ and $B$ equal to zero. In that case, given the properties of the Bessel function $J_{n}(0)=0$ for $n\ne\,0$ and $J_0(0)=1$, only the first line of the expression for $T_{L,L',\ell}$ (see APPENDIX) is non-zero, and after integration, the unaltered decay rate of the muon is recovered
\begin{equation}\label{G0}
\Gamma^0\,=\,\frac{G^2M^5}{192\pi^3}
\end{equation}
where, as in \cite{us}, terms proportional to the electron mass have been neglected.

As explained in our previous treatment \cite{us}, in \eqref{GLL'l} the limits of integration are determined by the $\Theta$ function, and the integration separates into two parts
\begin{equation}\label{I3}
\int\,dE\,\int\,dz\,\Theta(Q^2)\,=\,\int_{m}^{\frac{M}{2}}\,dE\,\int_{-1}^{1}\,dz
                                         +\int_{\frac{M}{2}}^{\frac{M}{2}+(2L + \ell)\omega}dE\,\int_{z_L(E)}^1\,dz
\end{equation}
with $z_L (E)$ coming from the the condition $Q^2 \ge 0$
\begin{equation}\label{z_L}
z_{L}(E)\,=\,-\frac{M^2+2M(2L + \ell)\,\omega-2E \, [M+(2L + \ell)\,\omega]}{2(2L + \ell)\,\omega\,E}\;.
\end{equation}
Since the first term of \eqref{I3} is independent of the summation indices, substitution of \eqref{I3} into \eqref{GLL'l} allows us to sum before integrating over this first term. In order to perform the summation per index of  $T_{L,L',\ell}$, we use $\sum_{n=-\infty}^{\infty}J_{n}(z)J_{n+k}(z)\,=\,J_k(0)$ (from {\bf 9.1.75} of \cite{AS}), the recurrsion relation for Bessel functions, $n\,J_{n}(z)\,=\,\frac{z}{2}(J_{n+1}(z)+J_{n-1}(z))$, and $J_{-n}(z)\,=\,(-1)^{n}J_{n}(z)$. We note that $(L-L')$ can only equal integer values; this notion forces many of the summation terms to be zero.
After some tedious calculation of the summations, the integration may be performed, resulting in (neglecting the electron mass terms)
\begin{equation}\label{gammaA}
\Gamma \,  = \, \Gamma^0 \, \Bigg \{ 1+ \frac{8e^2 a^2}{M^2} \brac{-\frac{5}{3} + \ln \frac{M}{m}} - \frac{e^4 a^4}{4M^4} \brac{6 - 10 \ln \frac{M}{m} + \frac{M^2}{m^2}} \Bigg \} \; .
\end{equation}
As we will see, the term proportional to $M^2 / m^2$ cancels with a similar term coming from the second part of the integration in \eqref{I3}.
\newpage
The second integration term can also be tackled in the same way described in \cite{us}. The indices of the Bessel functions are limited by $\ell \le D$, and $L \le B$, as Bessel functions become very small once the index exceeds the argument. Therefore, given typical values such as $ea \sim 10^{-4}$ MeV and $\omega \sim 1$ eV, $(2L + \ell)\omega$ is always much less than 1 MeV. Hence, the corrections from this integration term are small, as the range of integration is small. The integration can, thus, be expanded in a Taylor series, and after keeping only the first non-zero term, the correction becomes \cite{us}
\begin{equation}\label{gammaC}
\Gamma^C\,=\,\Gamma^0 \, \frac{4\omega}{M^5}\int_{-1}^1\,dz\sum_{L, L', \ell=-\infty}^{\infty}(2L + \ell) \,\widetilde{T}_{L,L',\ell}
\end{equation}
where $\widetilde{T}_{L,L',\ell}$ is the same as $T_{L,L',\ell}$ but without the prefactor of $128 \, G^2$. Note that $z_L(\frac{M}{2})=-1$ from \eqref{z_L}, and hence, once again we are allowed to switch the order of summation and integration. Although the calculation of \eqref{gammaC} is lengthy and extremely tedious, it results in a quite simple expression
\begin{equation}\label{gammaB}
\Gamma^C \,  = \, \Gamma^0 \, \Bigg \{ \frac{8e^2 a^2}{M^2} \brac{\frac{\omega^2}{M^2} - \frac{2 \omega^2}{M^2} \ln \frac{M}{m}} - \frac{e^4 a^4}{4M^4} \brac{3 - 2 \ln \frac{M}{m} - \frac{M^2}{m^2}} \Bigg \} \; .
\end{equation}
The total change of the decay rate is then the sum of \eqref{gammaA} and \eqref{gammaB}
\begin{equation}\label{gammaT}
\Gamma \,  = \, \Gamma^0 \, \Bigg \{ 1 + \frac{8e^2 a^2}{M^2} \brac{ \Big(1- \frac{2 \omega^2}{M^2} \Big) \ln \frac{M}{m} -\frac{5}{3} + \frac{\omega^2}{M^2}} - \frac{e^4 a^4}{4M^4} \brac{9 - 12 \ln \frac{M}{m}} \Bigg \} \; .
\end{equation}
Note the cancellation of the term proportional to $M^2/m^2$, once \eqref{gammaA} and \eqref{gammaB} are added.

\section{Discussion}

Comparing the final result \eqref{gammaT} with that for the circularly polarized laser,
\begin{equation}
\Gamma_{\rm circ}  
\,=\,\Gamma^0\Big\{1+8\frac{e^2a^2}{M^2}\Big[\big(\frac{\omega}{M}+4\frac{\omega^2} {M^2}\big)\ln\frac{M}{m}
+\frac{2}{3}-\frac{3}{2}\frac{\omega}{M}-2\frac{\omega^2}{M^2}\Big]\Big\}\,,
\end{equation}
there are some interesting differences. The corrections to the unperturbed decay rate are not the same in both polarization cases.  There are terms proportional to $e^4 a^4/M^4$ in a linearly polarized laser field and the correction in the linear case is larger than in the circular case because the $\ln(M/m)$ appears with a coefficient of 1 rather than a coefficient of $\omega/M$. Part of this difference is attributable to the fact that terms involving $\varepsilon^{\alpha\beta\gamma\delta}$ contribute to the circularly polarized case but do not contribute in the linearly polarized case.

Our calculations in the present paper, as well as in the previous one \cite{us}, show that the corrections to the undisturbed decay rate in both laser polarization cases are, however, consistently small, as one might  expect intuitively.

\section{Appendix}

The explicit expression for $T_{L,L',\ell}$, prior to the integration over the electron momentum and triple summation, and after simplification, is given by
\begin{align}
T_{L,L',\ell} = & \, 128 \, G^2 \times \notag \\
& \Bigg\{ J_L(B) J_{L'}(B) J_{\ell}(D) J_{2\brac{L-L'}+\ell}(D) \Big[ \brac{P-q}^2 \brac{P \cdot q} + 2 \brac{P-q} \cdot P \brac{P-q} \cdot q \notag \\
& \qquad \qquad \qquad \qquad \qquad \qquad \qquad \qquad + \frac{e^2 a^2}{4 q \cdot k} \big[ \brac{P-q}^2 \brac{P \cdot k} + 2 \brac{P-q} \cdot P \brac{P-q} \cdot k \big] \Big] \label{0} \displaybreak \\
& + \brac{2 \ell +4L} J_L(B) J_{L'}(B) J_{\ell}(D) J_{2\brac{L-L'}+\ell}(D) \Big[ \brac{P-q} \cdot q \brac{P \cdot k} + \brac{P-q} \cdot k \brac{P \cdot q} \notag \\
& \qquad \qquad \qquad \qquad \qquad \qquad \qquad \qquad \qquad \qquad \; + \brac{P-q} \cdot P \brac{q \cdot k} + \frac{e^2 a^2}{2 q \cdot k} \brac{P-q} \cdot k \brac{P \cdot k} \Big] \label{l0} \\
& +\brac{8 \ell L + 8 L^2 + 2 \ell^2} J_L(B) J_{L'}(B) J_{\ell}(D) J_{2\brac{L-L'}+\ell}(D) \Big[ \brac{P \cdot k} \brac{q \cdot k} \Big] \label{lL0} \\
& +J_L(B) J_{L'}(B) J_{\ell}(D) J_{2\brac{L-L'}+\ell+1}(D) \frac{e}{4 q \cdot k} \Big[ \big[ \brac{q\cdot a} \brac{P\cdot k} - \brac{P\cdot a} \brac{q\cdot k}  \notag \\
& \qquad \qquad \qquad \qquad \qquad \qquad \qquad \qquad \qquad \qquad  +i \, \varepsilon_{\alpha\beta\gamma\delta}k^{\alpha}q^{\beta}a^{\gamma}P^{\delta}) \big]\brac{3M^2-4P\cdot q+m^2} \Big] \label{1} \\
& + \brac{\ell + 2L} J_L(B) J_{L'}(B) J_{\ell}(D) J_{2\brac{L-L'}+\ell+1}(D) \frac{e}{2 q \cdot k} \Big[ \big[ \brac{q\cdot a} \brac{P\cdot k} - \brac{P\cdot a} \brac{q\cdot k}  \notag \\
& \qquad \qquad \qquad \qquad \qquad \qquad \qquad \qquad \qquad \qquad\qquad\qquad +i \, \varepsilon_{\alpha\beta\gamma\delta}k^{\alpha}q^{\beta}a^{\gamma}P^{\delta}) \big]\brac{2P\cdot k-q\cdot k} \Big] \label{l1} \\
& + J_L(B) J_{L'}(B) J_{\ell}(D) J_{2\brac{L-L'}+\ell-1}(D) \frac{e}{4 q \cdot k} \Big[ \text{bracket} \eqref{1} \Big] \label{-1} \\
& +\brac{\ell + 2L} J_L(B) J_{L'}(B) J_{\ell}(D) J_{2\brac{L-L'}+\ell-1}(D) \frac{e}{2 q \cdot k} \Big[ \text{bracket} \eqref{l1} \Big] \label{l-1} \\
& +J_L(B) J_{L'}(B) J_{\ell+1}(D) J_{2\brac{L-L'}+\ell}(D) \frac{e}{4 q \cdot k} \Big[ \big[ \brac{q\cdot a} \brac{P\cdot k} - \brac{P\cdot a} \brac{q\cdot k}  \notag \\
& \qquad \qquad \qquad \qquad \qquad \qquad \qquad \qquad \qquad \qquad -i \, \varepsilon_{\alpha\beta\gamma\delta}k^{\alpha}q^{\beta}a^{\gamma}P^{\delta}) \big]\brac{3M^2-4P\cdot q+m^2} \Big] \label{p1}  \\
& + \brac{\ell + 2L} J_L(B) J_{L'}(B) J_{\ell+1}(D) J_{2\brac{L-L'}+\ell}(D) \frac{e}{2 q \cdot k} \Big[ \big[ \brac{q\cdot a} \brac{P\cdot k} - \brac{P\cdot a} \brac{q\cdot k}  \notag \\
& \qquad \qquad \qquad \qquad \qquad \qquad \qquad \qquad \qquad \qquad\qquad\qquad -i \, \varepsilon_{\alpha\beta\gamma\delta}k^{\alpha}q^{\beta}a^{\gamma}P^{\delta}) \big]\brac{2P\cdot k-q\cdot k} \Big] \label{lp1} \\
& + J_L(B) J_{L'}(B) J_{\ell-1}(D) J_{2\brac{L-L'}+\ell}(D) \frac{e}{4 q \cdot k} \Big[ \text{bracket} \eqref{p1} \Big] \label{p-1}  \\
& +\brac{\ell + 2L} J_L(B) J_{L'}(B) J_{\ell-1}(D) J_{2\brac{L-L'}+\ell}(D) \frac{e}{2 q \cdot k} \Big[ \text{bracket} \eqref{lp1} \Big] \label{lp-1} \\
& - J_L(B) J_{L'}(B) J_{\ell+1}(D) J_{2\brac{L-L'}+\ell+1}(D) \frac{e^2 a^2}{8 q \cdot k} \Big[ \brac{P-q}^2 \brac{P \cdot k} + 2 \brac{P-q} \cdot P \brac{P-q} \cdot k \Big] \label{p11} \raisetag{20pt} \\
& - \brac{\ell + 2L} J_L(B) J_{L'}(B) J_{\ell+1}(D) J_{2\brac{L-L'}+\ell+1}(D) \frac{e^2 a^2}{2 q \cdot k} \Big[ \brac{P-q} \cdot k \brac{P \cdot k} \Big] \label{lp11} \\
& - J_L(B) J_{L'}(B) J_{\ell+1}(D) J_{2\brac{L-L'}+\ell-1}(D) \frac{e^2 a^2}{8q \cdot k}\Big[ \brac{P-q}^2 \brac{P \cdot k} + 2 \brac{P-q} \cdot P \brac{P-q} \cdot k \Big]  \label{p1-1} \\
& - \brac{\ell + 2L} J_L(B) J_{L'}(B) J_{\ell+1}(D) J_{2\brac{L-L'}+\ell-1}(D) \frac{e^2 a^2}{2q \cdot k}\Big[ \brac{P-q} \cdot k \brac{P \cdot k} \Big] \label{lp1-1} \\
& - J_L(B) J_{L'}(B) J_{\ell-1}(D) J_{2\brac{L-L'}+\ell+1}(D) \frac{e^2 a^2}{8q \cdot k}\Big[ \brac{P-q}^2 \brac{P \cdot k} + 2 \brac{P-q} \cdot P \brac{P-q} \cdot k \Big]  \label{p-11} \\
& - \brac{\ell + 2L} J_L(B) J_{L'}(B) J_{\ell-1}(D) J_{2\brac{L-L'}+\ell+1}(D) \frac{e^2 a^2}{2q \cdot k}\Big[ \brac{P-q} \cdot k \brac{P \cdot k} \Big]  \label{lp-11} \\
& - J_L(B) J_{L'}(B) J_{\ell-1}(D) J_{2\brac{L-L'}+\ell-1}(D) \frac{e^2 a^2}{8q \cdot k}\Big[ \brac{P-q}^2 \brac{P \cdot k} + 2 \brac{P-q} \cdot P \brac{P-q} \cdot k \Big]  \label{p-1-1} \\
& - \brac{\ell + 2L} J_L(B) J_{L'}(B) J_{\ell-1}(D) J_{2\brac{L-L'}+\ell-1}(D) \frac{e^2 a^2}{2q \cdot k}\Big[ \brac{P-q} \cdot k \brac{P \cdot k} \Big] \label{lp-1-1} \Bigg\}
\end{align}
where ``bracket(*)" means the expression in the bracket of the corresponding ``*"--equation, and $\varepsilon_{0123} = 1$.

\acknowledgments
DAD was supported in part by the U. S. Department of Energy under grant No. DE-FG03-93ER40757.
WWR was supported in part by the National Science Foundation under Grant PHY-0555544.

\end{document}